\documentclass[twocolumn]{autart}    %
\usepackage{graphicx}          %
\usepackage{amsmath}
\usepackage{dsfont}
\usepackage{amssymb}
\usepackage{siunitx}
\usepackage{caption}
\usepackage{xcolor}
\DeclareMathAlphabet{\mathpzc}{OT1}{pzc}{m}{it}

\newcommand\thefont{\expandafter\string\the\font}
\newcommand{\figref}[1]{\figurename~\ref{#1}}

\newcommand{\pdff}[0]{p^\mathrm{F}} %
\newcommand{\pdfh}[0]{p^\mathrm{H}} %
\newcommand{\pY}{p^\mathrm{y}}     %
\newcommand{\pX}{p^\mathrm{x}}     %
\newcommand{\Minv}{M^{\dagger}}
\newcommand{\nnormals}{n_\mathrm{p}}   %
\newcommand{\nz}{n_\mathrm{z}}
\newcommand{\pipe}{\hspace{0.2mm}|\hspace{0.2mm}}

\newcommand{\TR}[1]{#1}
\newcommand{\change}[1]{#1}

\makeatletter
\g@addto@macro\normalsize{%
  \setlength\abovedisplayskip{7.5pt}
  \setlength\belowdisplayskip{7.5pt}
  \setlength\abovedisplayshortskip{7.5pt}
  \setlength\belowdisplayshortskip{7.5pt}
}

\begin{document}

\begin{frontmatter}

\runtitle{Meta-State-Space Learning: An Identification Approach for Stochastic Dynamical Systems} %
\title{Meta-State-Space Learning: An Identification Approach for Stochastic Dynamical Systems}

\author[Eindhoven]{Gerben I. Beintema}\ead{g.i.beintema@tue.nl},    %
\author[Eindhoven]{Maarten Schoukens}\ead{m.schoukens@tue.nl},               %
\author[Eindhoven,Budapest]{Roland Toth}\ead{r.toth@tue.nl}  %
\thanks[footnoteinfo]{Implementation of the proposed approach is available in the toolbox: {\tt github.com/GerbenBeintema/metaSI}, while the software for the examples considered in this paper can be found at: {\tt github.com/GerbenBeintema/metaSS-code-repo}. 
}

\thanks[footnoteinfo2]{This research has been funded by the European Union (ERC, COMPLETE, 101075836) and also by the Eötvös Loránd Research Network (Grant Number: SA-77/2021). Views and opinions expressed are however those of the author(s) only and do not necessarily reflect those of the European Union or the European Research Council Executive Agency. Neither the European Union nor the granting authority can be held responsible for them.}

\address[Eindhoven]{Department of Electrical Engineering, Eindhoven University of Technology, %
Eindhoven, The Netherlands}  %
\address[Budapest]{Systems and Control Laboratory, HUN-REN Institute for Computer Science and Control, %
Budapest, Hungary.}             %


\begin{abstract}                          %
Available methods for identification of stochastic dynamical systems from input-output data generally impose restricting structural assumptions on either the noise structure in the data-generating system or the possible state probability distributions. In this paper, we introduce a novel identification method of such systems, which results in a dynamical model that is able to produce the time-varying output distribution accurately without taking restrictive assumptions on the data-generating process. The method is formulated by first deriving a novel and exact representation of a wide class of nonlinear stochastic systems in a so-called meta-state-space form, where the meta-state can be interpreted as a parameter vector of a state probability function space parameterization. As the resulting representation of the meta-state dynamics is deterministic, we can capture the stochastic system based on a deterministic model, which is highly attractive for identification. The meta-state-space representation often involves unknown and heavily nonlinear functions, hence, we propose an \emph{artificial neural network} (ANN)-based identification method capable of efficiently learning nonlinear meta-state-space models. We demonstrate that the proposed identification method can obtain models with a log-likelihood close to the theoretical limit even for highly nonlinear, highly stochastic systems.
\vspace{-3mm}
\end{abstract} 
\end{frontmatter}

\section{Introduction}\label{sec:intro}

The framework of stochastic dynamical systems and (hidden) Markov models is a well-respected field of study with many powerful results and application areas~\cite{paul1999stochastic}. \change{ \TR{From model predictive control with uncertain stochastic dynamics and probabilistic constraints \cite{7740982}, till filtering with Kalman variants up to particle filters \cite{pfilter}, %
general stochastic models have been extensively applied in many practical applications from the process industry to aerospace engineering.}
\TR{As often the} relations governing the system behavior are \TR{only partly known in practice},} %
estimation of a reliable model of the dynamics from measured \emph{input-output} (IO) data has been a central question of research~\cite{paul1999stochastic}.

\change{\TR{For the case, when the dynamics are}} \emph{Linear-time-invariant} (LTI), identification algorithms \change{\TR{have been developed}} to accomplish this by estimating both a process model and a noise model. Examples of such methods include \change{\emph{subspace approaches}~\cite{van1994n4sid,larimore1990CVA,verhaegen1992MOESP} and \emph{prediction error methods,} resulting in IO models with \emph{autoregressive with exogenous input} (ARX), \emph{moving average ARX} (ARMAX), \emph{output-error} (OE), and \emph{Box-Jenkins} (BJ) noise structures} or state-space models, e.g., with an \emph{innovation noise} structure~\cite{ljung1999SI}. \change{In comparison, most \emph{nonlinear-time-invariant} (NLTI) identification algorithms are highly restrictive on the noise structures which are allowed to be present in the system. The noise models are} mostly limited to \change{\emph{nonlinear ARX} (NARX) and \emph{nonlinear OE} (NOE)} models in the input-output representation case~\cite{schoukens2019nonlinear}, and innovation noise structures in the state-space case~\cite{van2009subspace}. These noise models are commonly considered as one of their advantages is that they allow estimation of the noise corrupting the data, under the assumption that the data-generating system itself falls into these model classes. When considering highly-structured models, such as block-oriented models, specific structured nonlinear noise models have been also considered~\cite{Schoukens2017,Schoukens2012,Hagenblad2008}. 

Kernel-based methods such as regularization networks, support vector machines, and Gaussian regression offer another class of system identification approaches, both in the LTI and NLTI cases, with a robust mathematical framework and direct estimation of model uncertainty~\cite{chiuso2019kernelmethods}. Such methods have been applied in system identification under various noise assumptions, e.g., white process noise under full-state measurement~\cite{eleftheriadis2017GP-SS,deisenroth2011pilco}, innovation noise~\cite{shakib2020kernel-inno}, and equation-error noise \cite{Pillonetto:2011}. Kernel methods can theoretically be extended to most model and process noise structures by appropriate selection of the involved kernels, however, currently, there exists no general systematic approach for appropriate kernel construction \TR{when the %
noise dynamics are unknown}%
~\cite{chiuso2019kernelmethods}. While conceptually kernel selection could be automatized with for instance genetic programming~\cite{khandelwal2023genetic}, the involved computational effort could be overwhelming for many practical identification problems. 

Another method is the identification of probabilistic state-space models through expectation maximization~\cite{schon2011system} using particle and smoothing filters. This method does not impose major restrictions on either the noise or the model structure, however, it can have a significant computational cost due to the Monte Carlo nature of such particle-based approaches. 

\change{In this paper, we contribute to resolving the current challenges in stochastic dynamical system identification by proposing \TR{a} so-called meta-state-space representation \TR{of the dynamics} and \TR{an} identification algorithm \TR{to estimate it from data}. In contrast to previous work, the meta-state-space \TR{representation} is directly applicable to a wide \TR{range} of nonlinear stochastic systems %
\TR{and it provides an}
exact representation where the meta-state can be \TR{seen as a} %
state probability function space parameterization.} An attractive property of the meta-state is that it fully represents the complete distribution of the original state, but its evolution, i.e., the meta-state transition function, is deterministic. Remarkably, the latter allows to capture of the stochastic process representation via a deterministic model, capable of describing the evolution of the complete \emph{probability density function} (PDF) of the \change{state} trajectories as a response to an input sequence. \change{Furthermore, since the output PDF is a function of the state PDF, the meta-state-space representation is also able to describe the \TR{PDF of the} output trajectories.} As the meta-state-space representation of a stochastic process often involves unknown and heavily nonlinear functions, we propose an \emph{artificial neural network} (ANN)-based identification method that, by exploiting the universal approximator capabilities of ANNs, is capable of learning such meta-state-space models efficiently directly from data. This provides a general approach for data-driven modelling of stochastic systems \TR{well beyond the capabilities of the current state-of-the-art} without \TR{any} severe structural restriction or limiting assumption. 

\change{To summarize,} the main contributions of the paper are:
\begin{itemize}
\item Showing that a wide class of nonlinear stochastic systems have a meta-state-space representation;
\item \change{Formulation of a stochastic system identification algorithm based on meta-state-space models and using only measured IO data;}
\item Solving the identification problem by a computationally efficient ANN-based \change{parameterization and} estimation approach.
\end{itemize}

This paper is structured as follows, Section \ref{sec:meta-state-derivation} introduces and proves the existence of the meta-state-space representation. In Section \ref{sec:meta-state-space-modelling}, we formulate the identification problem of stochastic systems via meta-state-space models using only IO data and propose a solution to it by an ANN-based approach. This is followed by Section \ref{sec:exp}, where capabilities of the proposed identification method are demonstrated on a challenging stochastic nonlinear system identification problem where the resulting estimation performance is found to be close to the theoretical limit. Lastly, conclusions on the achieved results and future research directions are provided in Section \ref{sec:conclusion}.

\section{The meta-state-space representation} \label{sec:meta-state-derivation}

Consider a discrete-time nonlinear stochastic system with process and measurement noise described by 
\begin{subequations}
\label{eq:NL-SS-sys}
\begin{align}
    x_{t+1} &= f_\mathrm{x}(x_t, u_t, v_t), \qquad y_t = h_\mathrm{x}(x_t, u_t, e_t),
\end{align}
\end{subequations}
where $x_t$ represents the state which is a random variable taking values from $\mathbb{X} \subseteq \mathbb{R}^{n_\mathrm{x}}$ with initial condition $x_0$ described by the PDF $p^\mathrm{x}_0:\mathbb{X}\rightarrow \mathbb{R}^+$, $u_t \in \TR{\mathbb{U}\subseteq  \mathbb{R}^{n_\mathrm{u}}}$ is a known input signal for simplicity of derivation (deterministic sequence or sample-path realisation of an input process), $y_t$ represents the output which is a random variable taking values from $\mathbb{Y} \subseteq \mathbb{R}^{n_\mathrm{y}}$ and $t\in\mathbb{Z}_0^+$ is the \TR{discrete} time. The output is corrupted by some i.i.d. stationary measurement noise $e$ with PDF $p^\mathrm{e}:\mathbb{R}^{n_\mathrm{e}}\rightarrow \mathbb{R}^+$. Furthermore, the state transition is also corrupted by some i.i.d. stationary process noise $v$ with PDF $p^\mathrm{v}:\mathbb{R}^{n_\mathrm{v}}\rightarrow \mathbb{R}^+$. Both $e$ and $v$ are considered to be independent of $u$. Lastly, $f_\mathrm{x}:\mathbb{X}\times\mathbb{U} \times \mathbb{R}^{n_\mathrm{v}} \rightarrow \mathbb{X} $ and $h_\mathrm{x}:\mathbb{X}\times\mathbb{U} \times \mathbb{R}^{n_\mathrm{e}} \rightarrow \mathbb{Y}$ are bounded functions of the state-transition and output functions respectively. 

The system described by the \emph{state-space} (SS) representation~\eqref{eq:NL-SS-sys} can be equivalently represented in the form of state-transition probabilities and conditional output probabilities, i.e., a \emph{hidden Markov model} or \emph{probabilistic state-space representation}: 
\begin{subequations}
\label{eq:conditional-probability-ss}
\begin{align}
    \pdff (x_{t+1}  \pipe x_t, u_t) &= \smallint p(x_{t+1}  \pipe x_t, u_t, v_t) p(v_t) \hspace{0.3mm} dv_t,  \\
     &= \smallint  \delta (x_{t+1} \! -\! f_\mathrm{x}(x_t, u_t, v_t)) p(v_t) \hspace{0.3mm} dv_t, \nonumber \\
    \pdfh(y_t  \pipe x_t, u_t) &= \smallint  p (y_t \pipe x_t, u_t, e_t) p(e_t) \hspace{0.3mm}de_t,\\
     &= \smallint  \delta (y_t \!-\! h_\mathrm{x}(x_t, u_t, e_t)) p(e_t) \hspace{0.3mm}de_t, \nonumber
\end{align}
\end{subequations}
where $\delta$ is the \emph{Dirac delta} function and $p$ denotes the corresponding PDFs.
For clarity of the derivation, we will adapt the following notation to indicate state and output probabilities at time $t\in\mathbb{Z}_0^+$: 
$$
\pX_t(x) \triangleq p(x_t),\ \ \ \ \ \pY_t(y) \triangleq p(y_t).
$$
This notation allows us to express all future probability distributions $\pX_t(x)$ and $\pY_t(y)$ given an initial state distribution $\pX_0(x) = p(x_0)$ and input signal using the \emph{Chapman--Kolmogorov equations}~\cite{paul1999stochastic}:  
\begin{subequations}  \label{eq:pdf-propagation-old}
\begin{align}
    \pX_{t+1}(x) &=  \int \pdff (x \pipe x', u_t) \pX_{t}(x') \hspace{0.3mm}dx', \\
    \pY_t(y) &= \int \pdfh(y \pipe x', u_t) \pX_t(x')\hspace{0.3mm} dx'.
\end{align}
\end{subequations}
We can also use functional operator notation to rewrite this in the following form
\begin{subequations}  \label{eq:pdf-propagation-new}
\begin{align}
    \pX_{t+1} &=  F(\pX_t, u_t), \label{eq:pdf-propagation-new:a}\\
    \pY_t &= H(\pX_t, u_t).
\end{align}
\end{subequations}
Introduce $u_{\tau}^{d}=[\begin{array}{cccc} u^\top\!(\tau) & u^\top\!(\tau+1) & \cdots & u^\top\!(\tau+d) \end{array}]^\top$. Using this notation and~\eqref{eq:pdf-propagation-new}, we can describe the state and output distribution evolution as
\begin{subequations}  \label{eq:F-recusive}
\begin{align}
    \pX_t &=  F^t(\pX_0,u_0^{t-1}), \\
    \pY_t &= H(F^t(\pX_0, u_0^{t-1}) , u_t),
\end{align}
\end{subequations}
where $F^t$ is defined in a recurrent manner as
\begin{subequations}
\begin{align}
    F^t(\pX_0,  u_0^{t-1}) &\triangleq F(F^{t-1}(\pX_0, u_0^{t-2}), u_{t-1}), \\
    F^0(\pX_0) &\triangleq \pX_0.
\end{align}
\end{subequations}
In terms of~\eqref{eq:F-recusive}, we can characterize all possible state distributions that can happen along the solution trajectories of~\eqref{eq:NL-SS-sys}:
\begin{multline} \label{eq:SX-def}
    \mathcal{S}^\mathbb{X} = \{ F^t(\pX_0, u_0^{t-1}) \hspace{0.3mm}|\hspace{0.3mm} \pX_0 \in \mathcal{S}^\mathbb{X}_0, u_0^{t-1} \in \mathbb{U}^t,  t\geq 0\}
\end{multline}
where $\mathcal{S}^\mathbb{X}_0$ are all initial state distributions of interest \TR{and  $\mathbb{U}^t  \subseteq \{ \{u_0, u_1,\ldots,u_t\} \mid u_i \in \mathbb{U} \subseteq \mathbb{R}^{n_\mathrm{u}}, t\geq i\geq0  \} $ is the set of all allowed input trajectories.}
Lastly, $\mathcal{S}^\mathbb{Y}$ is defined in a similar way for the output distributions. 

To derive the meta-state-space representation, we require that $\mathcal{S}^\mathbb{X}$ is parameterizable in terms of the following definition. 

\begin{defn}[Uniquely parameterizable PDF set] \label{def:par}
A set of probability density functions $\mathcal{S}^\mathbb{X}$ is called \emph{uniquely parameterizable of order $\nz$} if there exists an injective mapping $M : \mathcal{S}^\TR{\mathbb{X}} \rightarrow \mathbb{R}^{\nz}$. Hence, the inverse $\Minv$ exists on the co-domain of $M$ given $\mathcal{S}^\TR{\mathbb{X}}$. 
\end{defn}

Now we have all the ingredients to show the existence of the \emph{Meta-State-Space} (MSS) representation by the following theorem:

\begin{thm}[Meta-state-space representation] \label{th:2}
Assume that the set of probability functions $\mathcal{S}^\mathbb{X}$ formed by~\eqref{eq:NL-SS-sys} is \textit{uniquely parameterizable of order $\nz$} according to Definition \ref{def:par}. Then, there exist $f_\mathrm{z} : \mathbb{R}^{\nz} \times \mathbb{R}^{n_\mathrm{u}} \rightarrow \mathbb{R}^{\nz}$, $h_\mathrm{z} : \mathbb{R}^{\nz} \times \mathbb{R}^{n_\mathrm{u}} \rightarrow \mathcal{S}^\mathbb{Y}$ and $z_0 \in \mathbb{R}^{\nz}$ such that
\begin{subequations}
\begin{gather}
z_{t+1} = f_\mathrm{z}(z_t, u_t), \\
\pX_t = \Minv(z_t), \\
\pY_t = h_\mathrm{z}(z_t, u_t),
\end{gather}    
\end{subequations}
for all $t \geq 0$, all $\pX_0 \in \mathcal{S}^\mathbb{X}_0$, and all $u \in \mathbb{U}^\infty$. \vspace{-5mm}
\end{thm} 
\begin{pf}
    We provide the proof by induction: \\[1mm]
    \emph{Initial condition:} $\pX_0 = \Minv(z_0)$ by setting $z_0 = M(\pX_0)$. \\[1mm]
    \emph{Induction step:} If $\pX_t = \Minv(z_t)$, then
    \begin{align*}
        \pX_t &= \Minv(z_t)\ \hfill &\text{(apply $F$)}\\
        F(\pX_t, u_t) &= F(\Minv(z_t), u_t)\ &\text{(use~\eqref{eq:pdf-propagation-new:a}}) \\
        \pX_{t+1} &= F(\Minv(z_t),u_t)\  &\text{(apply $\Minv M$ RHS)} \\
        \pX_{t+1} &= \Minv ( \underbrace{M( F(\Minv(z_t), u_t) )}_{f_\mathrm{z}(z_t, u_t)}) \hspace{-4cm}&
    \end{align*}
    and thus $\pX_{t+1} = \Minv(z_{t+1})$ holds with $$z_{t+1} = f_\mathrm{z}(z_t, u_t) \triangleq M(F(\Minv(z_t), u_t)).$$
    \emph{Output case:} 
    By applying $\Minv$: $$\pY_t = H(\pX_t, u_t) = H(\Minv(z_t), u_t) \triangleq h_\mathrm{z}(z_t, u_t).$$
    \vskip -8mm \hfill $\blacksquare$
\end{pf}

\begin{figure}[t]
    \centering
    \includegraphics[width=0.95\linewidth]{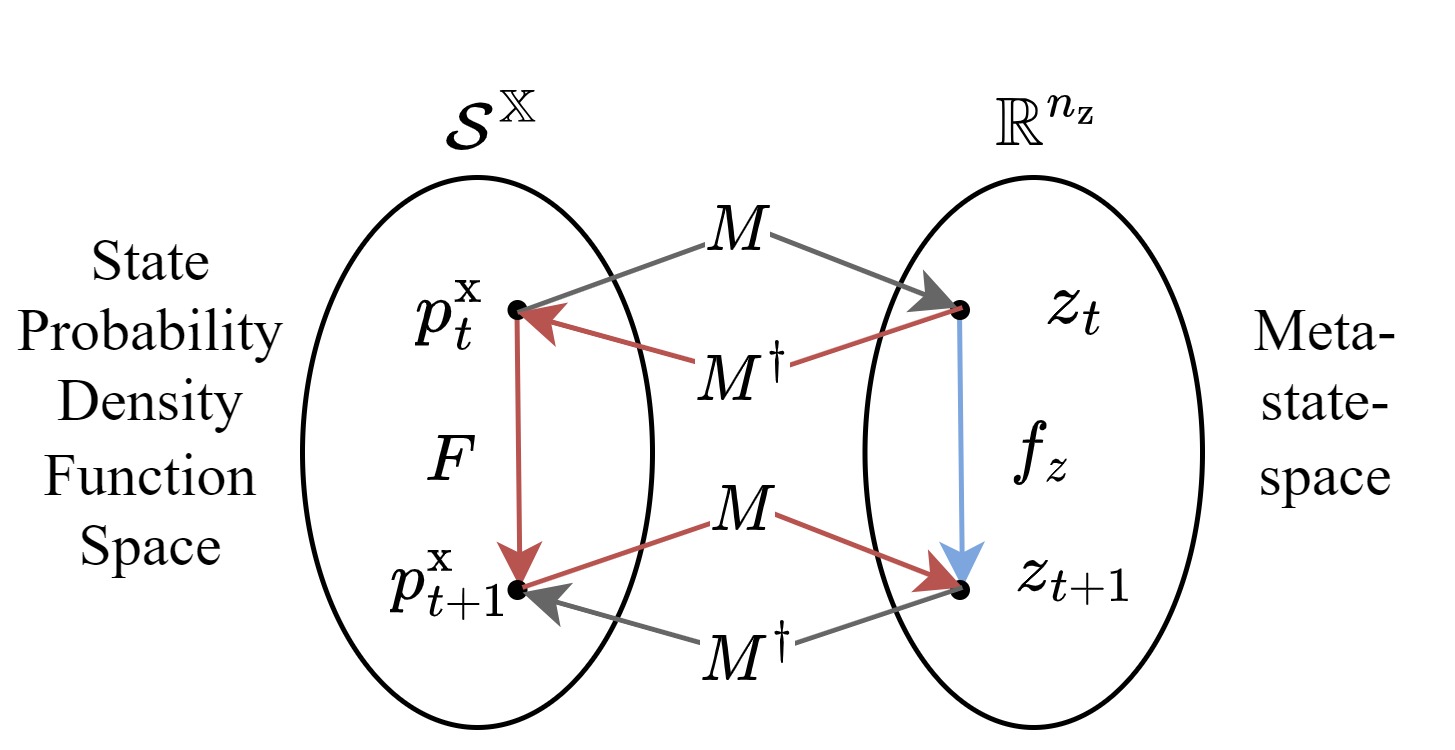}
    \caption{Both $\mathcal{S}^\mathbb{X}$ and the meta-state-space  with mapping $M$ and inverse mapping $\Minv$ \TR{are visualized, showing} that a transition from $z_t$ to $z_{t+1}$ can be \TR{computed} in two ways \TR{by} following either the blue or the red path and thus $z_{t+1} = f_\mathrm{z}(z_t,u_t) =  M(F(\Minv(z_t),u_t))$.}
    \label{fig:sketch-proof-meta-state}
\end{figure}

A way to understand this proof is by viewing \figref{fig:sketch-proof-meta-state} which shows that $z_{t+1} = f_\mathrm{z}(z_t,u_t) = M(F(\Minv(z_t),u_t))$ by the properties of set mappings.

\begin{figure}[t]
    \centering
    \includegraphics[width=0.95\linewidth]{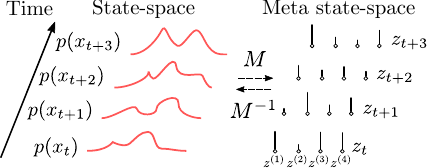}
    \caption{A graphical representation of the evolution of a PDF of the state according to~\eqref{eq:pdf-propagation-old} and the evolution of the meta-{state} in terms of~\eqref{eq:meta-state-space}. This figure shows that meta-state vectors $z_t$ can represent the state distribution $p_t^\mathrm{x}$ through the mapping $M$. }
    \label{fig:meta-state-overview} \vspace{-2mm}
\end{figure}

With this, we have shown the existence of an MSS representation of the system described by~\eqref{eq:NL-SS-sys} in the form:
\begin{subequations}
\label{eq:meta-state-space}
\begin{gather}
z_{t+1} = f_\mathrm{z}(z_t, u_t), \\
p(y_t \pipe z_t, u_t),
\end{gather} 
\end{subequations}
where $p(y_t \pipe z_t, u_t)$ is given by $h_\mathrm{z}(z_t, u_t)$. A graphical illustration of the evolution of the meta-state and its relation to the time-variation \TR{of the} original state distribution is given in \figref{fig:meta-state-overview}. The MSS representation is especially suited for system identification since~\eqref{eq:meta-state-space} is similar to the nonlinear SS representation of a deterministic system which has been studied extensively in the literature. 

\begin{rem} \label{rem:nz} Existence of an MSS, depends on the assumption that $\mathcal{S}^\mathbb{X}$ is \textit{uniquely parameterizable of order $\nz$} according to Definition \ref{def:par}. Hence it is an important question if such a parametrization exists or not for general nonlinear stochastic systems. It is well known that distributions in general can be uniquely defined in terms of their moments, which means that MSSs with potentially infinite order $\nz$ always exist. Higher-order moments have a diminishing role, and hence, often only a subset of the moments and thus finite $\nz$ is enough to provide an accurate characterisation of $\mathcal{S}^\mathbb{X}$. Additionally, there exist many universal approximators which can describe function spaces to arbitrary accuracy with increasing order $\nz$. For example, the difference becomes arbitrarily small with increasing the number of particles \cite{del1997particleacc} or increasing the number of components in a Gaussian mixture~\cite{Goodfellow2016deeplearningbook}. These approaches have been exploited in particle filtering~\cite{schon2011system} and Markov-chain Monte-Carlo methods~\cite{intro:chua2018PETS} to provide state-filtering and estimation for general stochastic systems. Hence the motivation for the existence of MSS models of~\eqref{eq:NL-SS-sys} with finite $\nz$ in an exact or approximative sense is based on the same considerations. However, characterisation of the minimal order $\nz$ of unique parameterizations of $\mathcal{S}^\mathbb{X}$ for a given~\eqref{eq:NL-SS-sys} and its boundedness are open questions and are outside the scope of the current paper.
\end{rem}

\section{Identification by meta-state-space learning}  \label{sec:meta-state-space-modelling}

In this section, we will exploit the existence of MSS representations of stochastic systems in the form of~\eqref{eq:NL-SS-sys} to formulate an efficient data-driven modelling approach of such systems within the meta-state-space setting by using \emph{maximum a posteriori} (MAP) estimation.

\subsection{The identification problem}

{\it Data set:} An input sequence $\{u_t\}_{t=1}^N$, either generated as the sampling of an input process or as a deterministic sequence, and unknown initial state $x_0$ are applied on the system given by~\eqref{eq:pdf-propagation-new} and an output realization $\{ y_t^\ast\}_{t=1}^N$ is recorded. These two sequences are used to create the input-\TR{output} dataset:
\begin{subequations}
\begin{align}
\mathcal{U}_N &= \{u_1, u_2, ..., u_N\},\\ \mathcal{Y}_N &= \{y_1^*, y_2^*, ..., y_N^*\}.
\end{align}
\end{subequations}
{\it Model structure:}  To identify~\eqref{eq:NL-SS-sys}, estimation of the meta-state dynamics~\eqref{eq:meta-state-space} in terms of a parameterized model 
\begin{subequations}\label{eq:meta-state-model}
\begin{gather}
    \hat{z}_{t+1} = f_\theta(\hat{z}_{t},u_t), \\
    p_\theta(\hat{y}_t \pipe \hat{z}_t,u_t), 
\end{gather}
\end{subequations}
is considered, where $f_\theta : \mathbb{R}^{\nz} \times \mathbb{R}^{n_\mathrm{u}} \rightarrow \mathbb{R}^{\nz}$ and $p_\theta$ is a conditional PDF, both parameterized in terms of $\theta\in\mathbb{R}^{n_\theta}$. Furthermore, the initial state $\hat{z}_1$ is also considered to be a parameter.

{\it Identification criterion:} Estimation of $\theta \in \Theta \subseteq \mathbb{R}^{n_\theta}$ and the initial state $\hat{z}_1\in\mathbb{R}^{\nz}$ are considered in terms of maximisation of the posteriori likelihood over $\Theta \times \mathbb{R}^{\nz}$. Since the currently considered formulation of the meta-state-space representation cannot express the joint distribution of the output, e.g., $p(y_1, y_2 \pipe u_1, u_2)$ we aim to maximize the product of the posteriori of each individual term as
\begin{align}
    [\theta, \hat{z}_1]_\textsc{MAP} &= \text{arg}\max_{\theta,\hat{z}_1} \hspace{0.2mm} p(\theta, \hat{z}_1, \mathcal{U}_N) \prod_{t=1}^{N} p_\theta(y_t^* \pipe \hat{z}_t, u_t) , \nonumber
\end{align}
where $p(\theta, \hat{z}_1, \mathcal{U}_N)$ is a user-specified prior distribution. By log and mean transformation, the MAP estimation problem of the parameters of a meta-state model can be posed as the following optimization problem:
\begin{align}
    \label{eq:main-optimization}
    \min_{\theta,\hat{z}_1} \quad & - \frac{1}{N} \sum_{t=1}^{N} \log( p_\theta(y_t^* \pipe \hat{z}_t,u_t)) - \frac{1}{N} \log (p(\theta,\hat{z}_1, \mathcal{U}_N)) \nonumber \\
    \textrm{s.t.} \quad & \hat{z}_{t+1} = f_\theta(\hat{z}_t, u_t), \quad \forall t\in \mathbb{I}_1^N,
\end{align}
where  $\mathbb{I}_1^N=[1,N]\subset \mathbb{Z}$. \change{Similar to classical system identification, minimizing a simulation cost can result in a good initial meta-state $\hat{z}_1$ since any error in it can result in a transient error which increases the cost~\cite{forgione2022learning}.} An important observation is that this optimization problem neither requires $M$ to be defined nor will result in an estimate of $M$ \change{which can be viewed as both an advantage and disadvantage. It is advantageous since the optimal choice of $M$ is generally unknown and thus would be challenging to choose a priori. However, it is also disadvantageous since after estimation,} it is unknown how the meta-state relates to the true hidden state distribution. An exception  is \TR{when} $x_t$ can be directly observed (i.e. $y_t = x_t$), \change{\TR{as in this case}} $p_\theta(x_t \pipe \hat{z}_t)$, which corresponds to $\Minv$, \change{\TR{is described by the model estimate}}.

\subsection{Neural meta-state-space estimator}  \label{sec:neural-meta-state}

This section aims to make the optimization problem given by~\eqref{eq:main-optimization} computationally tractable for gradient-descent optimization algorithms. Moreover, this section will introduce an efficient way of parameterizing $f_\theta$ and $p_\theta$ using neural networks for general modelling purposes. 

Meta-state-space models can be considered under various parameterizations of the meta-state transition function $f_\theta$ and output distribution $p_\theta(y \pipe z, u)$. However, these functions can be rather complicated and heavily nonlinear, hence parametrization by artificial neural networks is desirable \TR{due to} their expressiveness and favourable computational aspects~\cite{Goodfellow2016deeplearningbook}.
We parameterize the mapping $z_+ = f_\theta(z,u)$ as a fully connected feedforward neural network with a linear bypass and $n$ hidden layers which can be expressed recursively as:
\begin{subequations}\label{eq:feed-forward-with-linear-residual}
\begin{align}
    \xi^{(0)} &= [z^\top\ u^\top]^\top,\\
    \xi^{(i+1)} &= \phi \left (A^{(i)} \xi^{(i)} + b^{(i)} \right ), \\
    z_+ &= A^{(n)} \xi^{(n)} + A_{\text{lin}} \xi^{(0)} + b^{(n)},
\end{align}    
\end{subequations}
where $\xi^{(i)} \in \mathbb{R}^{n_\text{hidden}^{(i)}}$ are the hidden latent variables associated with the layers, $\{A_{\text{lin}},A^{(0)},b^{(0)},...,$ $A^{(n_\text{layers})},b^{(n_\text{layers})}\}$ are the real-valued network parameters with appropriate dimensions, and $\phi$ is a static nonlinear activation function which is applied element-wise. \change{\TR{A good standard choice is the} tanh activation function for $\phi$ since it is effective for many deep learning and system identification tasks~\cite{beintema2023deep}. However, other activation functions such as ReLU, Gaussian etc. \TR{and their combination for different layers} can be more effective depending on the problem \TR{at hand} (e.g., ReLu for piece-wise linear problems), see~\cite{Goodfellow2016deeplearningbook} for an overview.}

Regarding $p_\theta$, a flexible parameterization for non-Gaussian distributions is the mixture of Gaussian distributions~\cite{bishop1994mixture} based on which we consider
\begin{gather} \label{eq:guassian-mixture-NN}
    p_\theta(y \pipe \xi) = \sum_{i=1}^{\nnormals} w_{i,\theta}(\xi) \cdot \mathcal{N}\bigl(y \pipe \mu_{i,\theta}(\xi), \Sigma_{i,\theta}(\xi)\bigr),
\end{gather}
where, $\xi = [z^\top\ u^\top]^\top$, $\nnormals$ \TR{is} the number of Gaussian components, $w_{i,\theta}:\mathbb{R}^{\nz+n_\mathrm{u}} \rightarrow [0,1]$ with $\sum_{i=1}^{\nnormals} w_{i,\theta} = 1$,
$\mu_{i,\theta}:\mathbb{R}^{\nz+n_\mathrm{u}} \rightarrow\mathbb{R}^{n_\mathrm{y}}$, and $\Sigma_{i,\theta}:\mathbb{R}^{\nz+n_\mathrm{u}} \rightarrow\mathbb{S}^{n_\mathrm{y}}$, where $\mathbb{S}^{n_\mathrm{y}}$ is the set of symmetric, positive definite matrices in $\mathbb{R}^{n_\mathrm{y}\times n_\mathrm{y}}.$
The weight $w_j$, mean $\mu_j$ and covariance matrix $\Sigma_j$ functions are chosen as fully connected feedforward ANNs with a similar structure as $f_\theta$. \change{To improve computational effectiveness we utilize ANNs with $n_p$ outputs such that only three neural networks are required to parameterize the weights $w_\theta$, means $\mu_\theta$ and covariance terms $\Sigma_\theta$.} The validity of the probability distribution (i.e. $\int p_\theta(y  \hspace{0.2mm}|\xi) dy = 1$ and $p_\theta(y  \hspace{0.2mm}|\xi) \geq 0$, $\forall \xi$) is ensured by the given constraints and enforced by choosing appropriate activation functions on the last layer of each neural network as discussed in Appendix \ref{sec:par-normal}. This type of distribution parameterization is also called a \emph{Mixture Density Network}~\cite{bishop1994mixture}. \TR{The considered}  parameterization is sufficient to describe any PDF function under the limit of $\nnormals \rightarrow \infty$ since a weighted sum of normal distributions is a universal approximator~\cite{bishop1994mixture,Goodfellow2016deeplearningbook}. Using a Mixture Density Network is advantageous since the log-likelihood as in~\eqref{eq:main-optimization} can be computed in an efficient manner. Lastly, mitigation of floating point errors/instabilities is essential and is  also discussed in Appendix \ref{sec:par-normal}.

The computational cost and optimization stability of problem~\eqref{eq:main-optimization} can be greatly enhanced by adopting a multiple shooting formulation~\cite{bock1981multiple-shooting}. Following the multiple shooting approach of~\cite{beintema2023deep} for conventional nonlinear state-space estimation, we are able to reduce the computational complexity by using independent subsections of the available data. Based on these, we can recast~\eqref{eq:main-optimization} as the following optimization problem:
\begin{subequations}
\label{eq:trucated-zero-state-meta-state-optim}
\begin{align}
    \min_{\theta} \quad & -\!\!\! \sum_{t=1}^{N-T+1} \!\!\!\sum_{k=k_\text{burn}}^{T-1}\!\!\! \log( p_\theta(y_{t+k}^* \pipe z_{t+k \pipe t}, u_{t+k})) ,\\
    \textrm{s.t.} \quad & z_{t+k+1 \pipe t} = f_\theta(z_{t+k \pipe t}, u_{t+k}),\ \ {\forall k\in\mathbb{I}_0^{T-1}} \\
     \quad & \hspace{7.2mm} z_{t \pipe t} = 0, \ \  {\forall t\in \mathbb{I}_1^{N-T+1}} 
\end{align}
\end{subequations}
in which we also assume a uniform prior. 
\change{\TR{For simplicity of the implementation and computational feasibility, we set the initial state $z_{t|t}$ in each section to be fixed to zero in \eqref{eq:trucated-zero-state-meta-state-optim}.}
This choice might result in a mismatch with the optimal initial meta-states and thus a transient error can be present. By including a small burn time \TR{$k_\text{burn}$} the effect of this transient error is greatly reduced in the case of fading memory systems.} This formulation also allows for the use of powerful batch optimization such as Adam~\cite{kingma2014adam} by not summing over all possible $t$. 

The proposed model structure and estimation method possess a number of important hyperparameters, \TR{which can be chosen based on the following} %
guidelines: 
\begin{itemize}
    \item A key hyperparameter is the order of the meta-state-space model $\nz$. As mentioned in Remark \ref{rem:nz}, $\nz$ should be chosen such that Definition \ref{def:par} is satisfied with the desired level of user-defined accuracy with, for instance, cross-validation. 
    \item The number of Gaussian components $\nnormals$ should be chosen in a similar manner, but our observations suggest that $\nnormals=20$ \TR{has been a} sufficient \TR{baseline choice} for all the datasets that we have considered.
    \item The $k_\text{burn}$ and $T$ can be chosen using n-step-error figures as described in~\cite{beintema2023deep}. Hence, $k_\text{burn}$ should be chosen larger than the transient observed in the n-step-error figure and $T$ a few times that transient length for stable systems. 
\end{itemize}

Other well-known modelling guidelines, for instance described in~\cite{beintema2023deep}, still hold. 

\section{Simulation studies}\label{sec:exp}

To demonstrate the capabilities of the proposed method, consider the following nonlinear stochastic system
\begin{subequations}
\label{eq:data-gen-system}
\begin{align}
    x_{t+1} &= \alpha(x_t, e_t) x_t + u_t,\\
    y_t &= x_t,
    \end{align}
\end{subequations}
where $\alpha(x_t, e_t) = 0.3+0.7 e^{- (x_t+e_t)^2}$ which satisfies $ \pipe \alpha(x_t, e_t) \pipe \leq 1$ to ensure stability. The process noise $e_t$ is i.i.d with uniform distribution $p(e_t) = \mathcal{U}(e_t  \pipe -0.5, 0.5)$.

We generate three separate datasets for training, validation and testing, employing a white input sampled from a zero-mean normal distribution with a standard deviation of 2 (i.e. $u_t \sim \mathcal{N}(u_t \pipe 0, 2)$). The training and validation set consists of 300k and 10k sample points respectively with $x_0 = 0$ as the initial state. The test set consists of 5000 trajectories of 4100 samples each, where the first 100 samples are discarded to exclude transient effects. These test trajectories all use the same input realization $u_t$, but different noise realizations $e_t$ from the considered distribution. These trajectories allow us to compare the model and system output distributions. 

\begin{table}[t]
\caption{\change{The MSS model, cost and optimization parameters.}}
\label{tab:MSS-parameters}
\begin{tabular}{c|c|c|c|c|c|c|c}
$n_\text{layers}$ & $n_\text{hidden}$ & $\nz$ & $n_p$ & $k_\text{burn}$ & $T$ & \begin{tabular}[c]{@{}c@{}}learn\\ rate\end{tabular} & \begin{tabular}[c]{@{}c@{}}batch\\ size\end{tabular} \\ \hline
2                 & 64                & 3     & 30    & 10              & 30  & $10^{-3}$                                            & 2048                                                
\end{tabular}
\end{table}

\change{The MSS model, cost and optimization parameters can be viewed in Table. \ref{tab:MSS-parameters} which shows that $f_\theta$ \TR{is} parameterized} with a 2 hidden layer neural network with 64 nodes per layer with tanh activation functions as described in~\eqref{eq:feed-forward-with-linear-residual}. To parameterize $p_\theta$, we utilize a Gaussian mixture model as in~\eqref{eq:guassian-mixture-NN} where $w_{.,\theta}$, $\mu_{.,\theta}$ and $\sigma_{.,\theta}$ are neural networks with the same structure as $f_\theta$. To train the meta-state-space model, we utilize the multiple-shooting-based loss function given by~\eqref{eq:trucated-zero-state-meta-state-optim} since it scales well to the large training dataset. We use the Adam optimizer and the following hyper-parameters, input-output normalization, early stopping using the validation set, $k_\text{burn} = 10$, $T = 30$, $\nnormals = 30$, batch size of 2048 and a learning rate of $10^{-3}$. Lastly, using cross-validation we found that the model accuracy expressed in mean log-likelihoods is 1.525, 1.674 and 1.678 for $\nz = 2, 3$ and $4$ respectively. Hence, increasing the meta-state dimension beyond $\nz = 3$ does not provide significant increase in model accuracy and thus $\nz = 3$ is chosen.

\begin{figure}[t]
    \centering
    \includegraphics[width=1\linewidth]{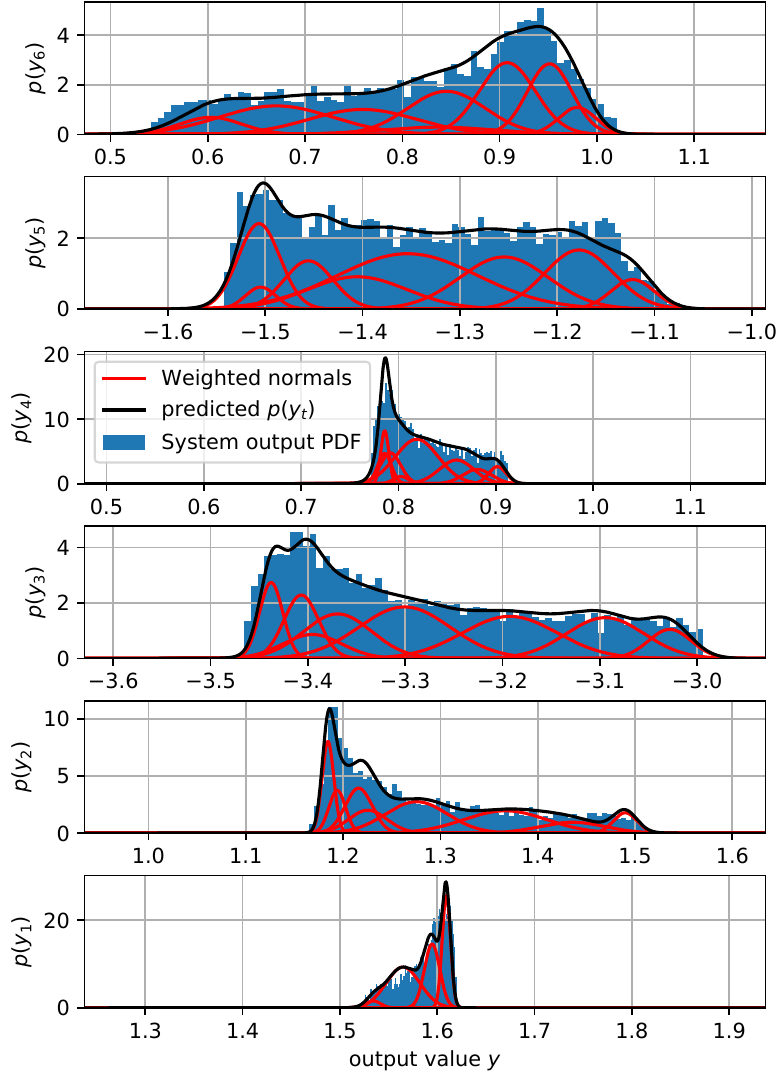}
    \caption{Probability density histograms of the system output under different noise realizations compared with the predicted output distribution of the meta-state-space model. Since the model output distribution is parameterized as a weighted sum of normal distributions, these weighted components are also displayed.}
    \label{fig:hist-and-y-dists}
\end{figure}

The resulting meta-state model is analyzed by using both qualitative and quantitative comparisons. The qualitative comparison aims to investigate if the produced probability distribution of the output trajectories of the model well represents the probability distribution of the output trajectories of the considered stochastic system. Using the 5000 test trajectories, we can construct a probability density histogram of the system output over time and compare it to the output distributions given by the model at specific time instances. This comparison can be viewed in \figref{fig:hist-and-y-dists} which shows a striking resemblance between the probability density histogram of the test set and the output distributions given by the model. Not only the mean and the variance have been captured by the model, but also smaller features such as \change{bumps as seen in the third row at $y=0.8$} are present in the histogram. With this, we have shown that the evolution of the meta-state $z_t$ can indeed describe the distributions of $y_t$ using a qualitative comparison. 

Numerically quantifying the model quality in a probabilistic setting is done by using the mean log-likelihood on the test set with $S=5000$ test sequences:
\begin{equation}\label{eq:mean-log-like}
    \frac{1}{N S} \sum_{t=1}^N \sum_{i=1}^S \log p(y_t^{*(i)} \pipe z_t)
\end{equation} 
where $(i)$ indicates the $i^{\text{th}}$ trajectory. To ease the interpretation of this quantity, we included a comparison of 3 baselines and an estimated theoretical upper limit. The following baselines are considered (see Appendix~\ref{sec:mean-var-defs} for details);
\begin{enumerate}
    \item Gaussian ($\mu_\mathrm{y}, \sigma_\mathrm{y}$): has a static mean and a static standard deviation. Comparison with this baseline gives an indication that the results are better than a static Gaussian model. 
    \item Gaussian ($\mu_{\mathrm{y},t}, \sigma_{\mathrm{y},\mathrm{s}}$): has a dynamic mean and a static standard deviation. This baseline represents a model which is able to perfectly capture the mean variation of the output, but assumes a static output noise. \change{The performance of this baseline is the upper limit of the performance of conventional output error modeling methods like~\cite{fraccaro2017classic-NL-dynamics}.} 
    \item Gaussian ($\mu_{\mathrm{y},t}, \sigma_{\mathrm{y},t}$); has a dynamic mean and a dynamic standard deviation. This baseline represents a model which is able to perfectly capture the mean variation  of the output, but assumes that the output noise is a state-dependent Gaussian. 
    \item Upper limit: it can be shown that for any model of the stochastic system, the mean output log-likelihood will be smaller or equal to the negative mean entropy of the output signal~\cite{vasicek1976entropy}, see Appendix~\ref{sec:noise-floor} for details. 
\end{enumerate}

The computed mean log-likelihood of all four baselines together with the obtained meta-state-space model are presented in Table~\ref{tab:results}. This table shows that the obtained model outperforms all three baselines and is very close to the estimated upper limit. This is a remarkable achievement since the meta-state-space model is able to describe the data generated by challenging stochastic dynamics using a deterministic state transition and a stochastic output map. Hence, the existence of the meta-state-space as derived in~\eqref{eq:meta-state-space} opens up efficient identification methods that directly estimate meta-state-space models. \change{Furthermore, we observe no major trends in the accuracy of the estimated model with increasing prediction horizon.} This identification method and possible future extensions can greatly reduce the complexity of the identification problem of stochastic dynamics. 

\newcommand{\doubleline}[2]{\begin{tabular}[c]{@{}l@{}}#1\\ #2\end{tabular}}

\begin{table}[t]
\caption{The mean model output log-likelihood~\eqref{eq:mean-log-like} over the test set for the nonlinear stochastic system given by~\eqref{eq:data-gen-system}.}
\label{tab:results}
\centering
\begin{tabular}{l|l}
Model/Baseline                       & \doubleline{Mean}{log-likelihood} \\ \hline
Gaussian ($\mu_\mathrm{y}, \sigma_\mathrm{y}$)         & -2.18     \\
Gaussian ($\mu_{\mathrm{y},t}, \sigma_{\mathrm{y},\mathrm{s}}$) & 1.04      \\
Gaussian ($\mu_{\mathrm{y},t}, \sigma_{\mathrm{y},t}$) & 1.56      \\
\textbf{Meta-state-space model}      & 1.67      \\
Upper limit          & 1.73     
\end{tabular}
\end{table}

\section{Conclusion}\label{sec:conclusion}
A novel meta-state-space identification method has been introduced which is able to identify \TR{general} nonlinear stochastic systems with an accuracy close to the theoretical limit as shown in the simulation study. The identification method is formulated based on a meta-state-space representation of the system which can be interpreted as a description of the deterministic evolution of a parameter vector of a state distribution parameterization, \TR{called} the meta-state. Identification based on this representation is effective since the meta-state transition function is deterministic \change{\TR{and in the considered example the proposed method could achieve accuracy close to the theoretical limit.} }

\change{\TR{By the current formulation, the estimated meta-state models allow}} only to express the output probability distributions of the type $p(y_t|u_0^t)$. However, we suspect that the meta-state can be extended \TR{for other prediction objectives}. For instance, by the inclusion of a subspace encoder~\cite{beintema2023deep} it is potentially possible to obtain accurate $n$-step ahead \TR{predictors such as} $p(y_{n+t}|u_0^{n+t}, y_0^n)$ which would be useful for model-based control with chance constraints. Furthermore, by the inclusion of a Kalman measure update in the meta-state-space, it is potentially possible to obtain the joint probability distributions $p(y_0^n|u_0^n)$ which would be useful for filtering and observer tasks. 

\bibliographystyle{plain}
\bibliography{references}

\appendix
\section{Mixture Density Network parametrization} \label{sec:par-normal}

Obtaining a valid mixture of Gaussians given by~\eqref{eq:guassian-mixture-NN} requires that the weights are $\sum_i w_i = 1$ and $w_i>0$ and that $\sigma_i >0$. This is enforced by utilizing proper activation functions as described below. Furthermore, floating-point errors are minimized for a successful implementation. 

To compute $w_i$ \TR{in \eqref{eq:guassian-mixture-NN}}, it is suggested to use the following relations
\begin{align}
    \tilde{w}_i &= \text{ANN}_{\theta w}^i(z), \\
    w_i &= \frac{\exp(\tilde{w}_i)}{\sum_j \exp(\tilde{w}_j)} =  \frac{\exp(\tilde{w}_i-\max_k (\tilde{w}_k))}{\sum_j \exp(\tilde{w}_j-\max_k (\tilde{w}_k))}.
\end{align}
\vskip -4mm \TR{Additionally, it is suggested to use} %
\begin{align}
     \tilde{\sigma_{i}} &= \text{ANN}_{\theta \sigma}^i(z), \\
    \sigma_{i} &= \exp(\tilde{\sigma_{i}}),\\
    \mu_i &= \text{ANN}_{\theta \mu}^i(z), 
\end{align}
where $n_y = 1$ with $\Sigma_i = \sigma_i^2$ \TR{is considered} for simplicity. Lastly, we compute the log probability as follows
\begin{align}
    r_i &\triangleq \log(w_i) + \log \left (N(y \pipe \mu_i, \sigma_i) \right),\\
    \log(p_\theta(y \pipe z)) &= \log \left ( \sum_i \exp(r_i) \right), \\
     = \max_k (r_k) &+ \log \left ( \sum_i \exp \left (r_i-\max_k (r_k)\right)\right).\nonumber
\end{align}
Here the $\max$ outside of the $\log$ reduces floating point errors which could prevent convergence of the optimization. Application of the $\max$-operator is well-known to improve numerical floating point stability, for instance, this has been the reason for the introduction of softmax activation functions in machine learning~\cite{karpathy2016cs231n}.  

\section{Mean log-likelihood upper limit} \label{sec:noise-floor}
It is well-known that the \emph{Kullback--Leibler} (KL) divergence is zero if the given distribution $q(y_t)$ is equal to the target distribution $p(y_t)$:
\begin{gather}
    D_\mathrm{KL}(p , q ) = \int p(y_t)  \log \left ( \frac{p(y_t)}{q(y_t)} \right ) dy_t,\\
     = \underbrace{\int p(y_t)  \log \left ( p(y_t) \right ) dy_t}_\text{negative differential entropy} -  \ \ \underbrace{\int p(y_t)  \log \left ( q(y_t) \right ) dy_t}_\text{cross entropy}, \nonumber
\end{gather}
where the cross entropy is equal to our performance measure using sampling of $p(y_t)$. This also shows that the upper bound is the negative mean entropy of $p(y_t)$. However, since an analytical expression for $p(y_t)$  is unavailable in practice, we cannot compute the differential entropy directly and will need to estimate it using the data samples. Many methods exist for computing entropy from samples. For this purpose, we employ the often applied method proposed by Vasicek, O. (1976) \cite{vasicek1976entropy,alizadeh2015entropy} to estimate the mean differential entropy. 

\section{Estimation of the Gaussian baselines} \label{sec:mean-var-defs}
Parameters of the Gaussian baseline models, discussed in Section \ref{sec:exp}, are obtained using the following equations:
\begin{align} 
     \mu_\mathrm{y} &= \tfrac{1}{N S} \Sigma_{t=1}^N \Sigma_{i=1}^S y_t^{*(i)}, \\
    \mu_{\mathrm{y},t} &= \tfrac{1}{S} \Sigma_{i=1}^S y_t^{*(i)}, \\
    \sigma_\mathrm{y}^2 &= \tfrac{1}{N S} \Sigma_{t=1}^N \Sigma_{i=1}^S (y_t^{*(i)} - \mu_\mathrm{y})^2, \\
    \sigma_{\mathrm{y},\mathrm{s}}^2 &= \tfrac{1}{N S} \Sigma_{t=1}^N \Sigma_{i=1}^S (y_t^{*(i)} - \mu_{\mathrm{y},t})^2, \\
    \sigma_{\mathrm{y},t}^2 &= \tfrac{1}{S} \Sigma_{i=1}^S (y_t^{*(i)} - \mu_{\mathrm{y},t})^2.
\end{align}

\end{document}